# Adaptive Dynamic Security for Operational Technology Against AI-Powered Attacks

Mohamed Chahine Ghanem

***Abstract*—Operational technology now faces attackers whose tempo is set by AI. We present ADS-OT, an adaptive dynamic security framework modelling defensive posture as a partially observable decision problem. A reinforcement-learning policy adapts monitoring, access, deception, and segmentation under a safety-governance layer, reducing attack-path completion, detection, and containment versus static baselines.**

## I. AI-Speed Attacks in Slow-Change Systems

On the plant floor, change is expensive. A controller that has run a packaging line for fifteen years talks to a SCADA server through a protocol that predates TLS and reboots once a year, during a planned outage. The defender's instinct is to keep things still: lock the configuration, audit the segmentation, revisit next quarter.

That instinct is now in tension with how attackers operate. Without needing fully autonomous malware, adversaries use large language models, code assistants, and reasoning agents to summarise vendor manuals, infer a site's likely engineering software, draft maintenance-themed phishing in the operator's language, vary payloads, and stitch weak signals into an attack path [13]. Reconnaissance that took weeks now takes an afternoon; phishing that read like a translation now reads like a colleague's note about a firmware update.

This is the tempo mismatch facing operational technology (OT). Programmable logic controllers (PLCs), engineering workstations, safety systems, remote-access gateways, and industrial Internet-of-Things (IIoT) devices have long lifetimes, fragile protocol stacks, strict availability needs, and deterministic traffic. Aggressive scanning can destabilise them; deep inspection adds latency; blocking a suspicious flow may halt production. Techniques unremarkable in enterprise networks — address hopping, workload migration, automatic isolation — violate operational assumptions where deterministic behaviour is itself a safety property.

Defenders must therefore decide dynamically: when to tighten posture, deepen monitoring, require stronger authentication, activate deception, or simply wait — each under uncertainty, where a process anomaly may be sensor drift or sabotage, a remote login overtime maintenance or a stolen credential, a command sequence normal in a change window and dangerous outside it [10].

M. C. Ghanem is with the School of Computer Science and Mathematics, Keele University, Newcastle-Under-Lyme, U.K. E-mail: m.c.ghanem@keele.ac.uk.

This article proposes *Adaptive Dynamic Security for Operational Technology* (ADS-OT), which models defence as a partially observable sequential decision problem: a reinforcement-learning (RL) policy operating over a constrained partially observable Markov decision process (POMDP) selects safe adaptations to monitoring, access control, deception, segmentation, protocol validation, and response readiness [7], [8]. It does not place an AI agent in command of industrial processes; it is a governed decision-support and controlled-automation layer wrapped in safety constraints, operator approval, auditability, and rollback.

We contribute an OT-specific threat-to-control framing (AI amplifies familiar attack stages rather than inventing new ones); a safety-constrained architecture in which a learning policy recommends posture changes while a governance layer decides what may execute; and an evaluation against three baselines on a small-to-medium enterprise (SME) manufacturing scenario, with cyber, operational, and governance metrics practitioners can carry into a cyber range or digital twin.

## II. What AI Adds to OT Attacks

AI does not create new categories of OT attack; it amplifies five behaviours that already mattered, documented in analyses of real OT threat-group tradecraft [11].

**Reconnaissance at scale.** Public documents, vendor manuals, job adverts, procurement notices, exposed banners, and forum posts reveal much about an industrial environment. A model can consolidate them into a profile of likely engineering tools, architectures, and target workflows before any intrusion; the attacker's first model of the plant may be more accurate than the defender's.

**Operationally plausible social engineering.** OT runs on specialist roles, vendors, maintenance windows, and urgent operational language. Generated messages imitate support tickets, firmware notices, and engineering handovers: phishing in this register is operationally framed, not merely personalised, and a message naming a plausible firmware version and maintenance task slips past scepticism far more reliably than generic phishing.

**Faster attack chaining.** A reused credential, an exposed remote-access appliance, a misconfigured firewall rule, and an insecure protocol may each be tolerable in isolation; together they form a path that AI tools find faster and with less skill.

**Evasion and variation.** Many malicious OT actions use legitimate tools and protocols [14]. The line between an engineer downloading new logic and an attacker doing the same lies in timing, authorisation, sequence, and context, not in a malicious binary. Code rewriting, payload variation, and log interpretation all become cheaper with AI assistance, sharpening living-off-the-land techniques.

**Learning the defender.** If a control always triggers at the same threshold, the threshold itself becomes intelligence. Models help an adversary probe what raises an alert, while defenders contend with high alert volumes and analyst fatigue [12]. A strong but predictable defence degrades over time; defenders need posture that is context-aware and adaptive, yet safety-constrained.

## III. From Static Controls to Adaptive Dynamic Security

Static OT security remains essential: inventory, secure configuration, segmentation, identity governance, remote access, backup, logging, patching, incident response, supplier assurance. NIST SP 800-82, NIST SP 800-53, IEC 62443, the NIST Cybersecurity Framework, and the CIS Controls supply the baseline vocabulary [1]–[5]; MITRE ATT&CK for Enterprise and ICS connects observed behaviour to adversary techniques [6]. None of this changes.

What changes is the tempo. Organisations pick controls, implement them, audit coverage, and update at intervals — necessary for governance, yet silent mid-campaign: should remote access now require step-up authentication; should protocol validation tighten around one workstation; should deception activate; is isolating a suspicious device riskier than the suspected compromise?

Adaptive dynamic security separates controls into two classes: *foundational* controls remain continuously active (inventory, baseline segmentation, secure configuration, backup, account governance, incident-response planning), while *elastic* mechanisms change intensity or priority (deeper logging, packet capture, session recording, deception, stricter authentication, protocol validation, temporary segmentation adjustments). Table I contrasts the two postures.

The point is not to keep the environment in motion — often the right action is to hold posture, and adaptation should be proportional, explainable, and reversible. ADS-OT extends selective defence [9], [15] from a static allocation problem to a dynamic policy one: how should defensive effort be reallocated as evidence changes, under hard safety constraints?

TABLE I
Static OT Security Versus Adaptive Dynamic Security

| Dimension | Static OT security | Adaptive dynamic security |
|---|---|---|
| Aim | Establish and audit baseline controls; no mid-campaign reprioritisation. | Adjust posture as risk, evidence, and operational context change. |
| Decision cycle | Periodic assessment and scheduled improvement; posture frozen between reviews. | Event-driven adaptation under explicit safety constraints. |
| Threat assumption | Patterns evolve slowly enough that review tempo matches attacker tempo. | Attackers use AI to accelerate reconnaissance, framing, evasion, and chaining. |
| Visibility | Alerts and assessments suffice; weak signals triaged in isolation, not fused. | Compromise and intent are partially observable and must be inferred. |
| Examples | Fixed segmentation, baseline logging, scheduled vulnerability management. | Dynamic monitoring intensity, adaptive authentication, deception, protocol validation, response staging. |
| Success measure | Coverage, compliance, auditability; silent on when and how strongly to adapt. | Risk reduction per unit of operational cost with no unacceptable safety impact. |

## IV. The ADS-OT Framework

Figure 1 illustrates the ADS-OT architecture: five interacting layers — knowledge preparation, belief estimation, a POMDP/RL decision engine, adaptive security mechanisms, and governance feedback. Crucially, it separates *recommendation* from *execution*: the decision engine proposes posture changes; the safety and governance layer decides whether, how, and under whose approval they apply.

The layers map one-to-one onto the figure: the telemetry band at its base supplies the knowledge-preparation inputs; the belief-estimation block fuses these into the belief state $b_t$; the decision engine consumes $b_t$ and proposes an adaptation from the adaptive-mechanisms band; and the safety-and-governance layer is deliberately drawn *between* engine and plant, because no proposal executes without passing its masking, compliance, approval, audit, and rollback checks. The figure's attack-surface reduction is the combined effect of three mechanism families: segmentation and service isolation shrink the reachable network surface, access control the credential surface, and deception absorbs reconnaissance so probes reveal the attacker rather than the plant.

### A. Knowledge Preparation

The framework begins with a knowledge base: asset inventories, zone and conduit models, communication dependencies, vulnerability and patch status, engineering and remote-access logs, process telemetry, threat intelligence, and ATT&CK mappings. Each asset risk carries four attributes:

$$\text{Risk} = \{\text{Value}, \text{Impact}, \text{AI} - \text{Amplification}, \text{Constraint}\}. \tag{1}$$

The AI-amplification field captures whether AI is likely to increase a vector's speed, scale, stealth, personalisation, or chaining potential: vendor-access phishing scores high because generative tools make messages operationally credible; an exposed remote-access appliance because automated chaining makes it likelier to be targeted.

### B. Belief Estimation

The defender rarely knows the true state of the plant. ADS-OT maintains a belief over hidden conditions — asset exposure, compromise likelihood, attacker stage, operational impact, telemetry confidence, defensive effectiveness — updated from noisy observations: authentication anomalies, unusual engineering-software activity, unexpected protocol commands, process deviations, operator reports. This lets weak signals combine: an unusual login alone is not a crisis, but the same login during a maintenance lull, from a host that has never run engineering software, against a privileged workstation, is.

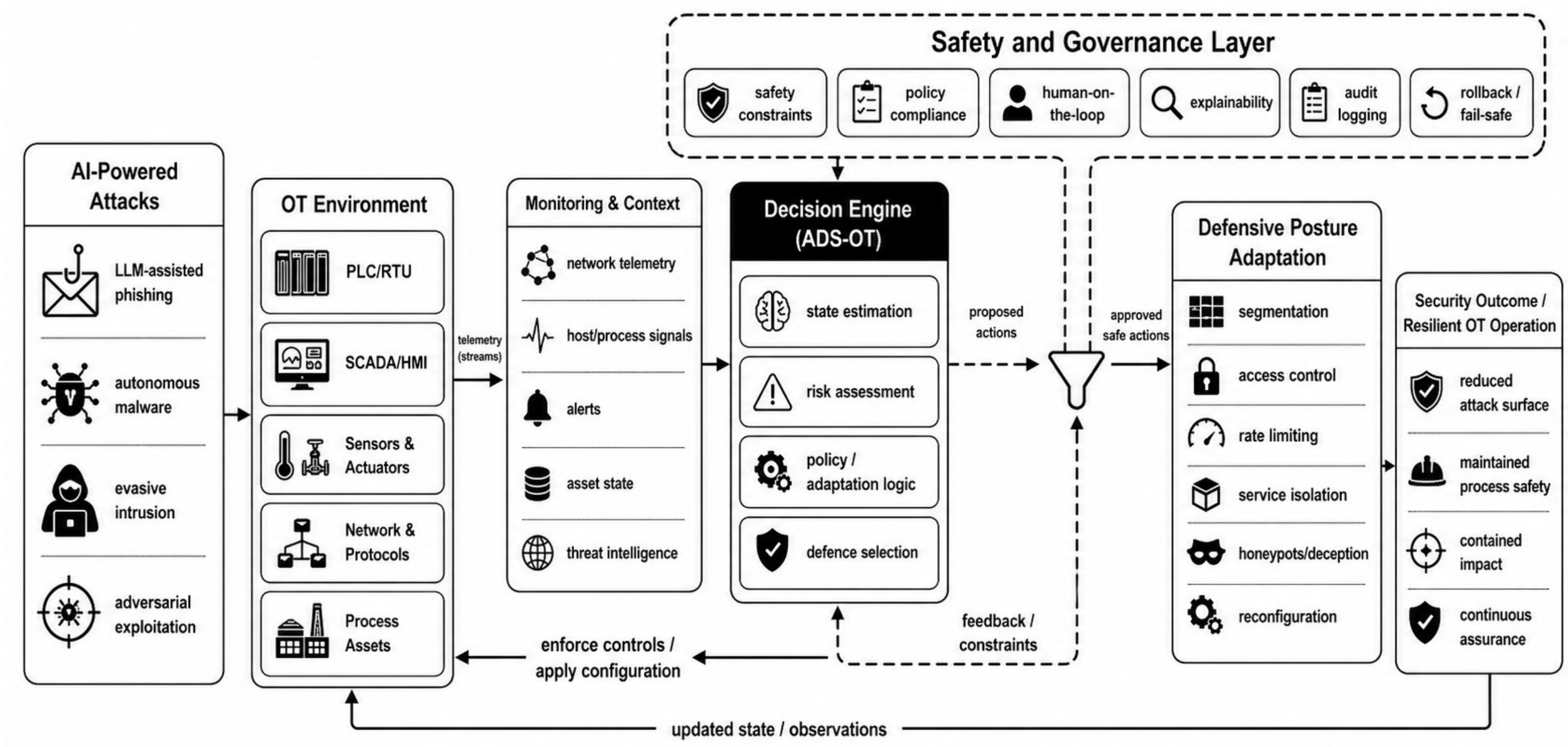

Fig. 1. The ADS-OT framework. Telemetry feeds a belief estimator; a POMDP/RL engine proposes adaptations; a safety and governance layer applies masking, compliance checks, human-on-the-loop approval, explainability, audit, and fail-safe rollback before any action reaches the plant.

### C. Decision Engine

The engine selects an adaptation conditioned on the current belief: what change reduces cyber-physical risk without unacceptable operational cost? Candidates include increased inspection, deception, step-up authentication, staged playbooks, a tightened conduit, or holding posture.

### D. Adaptive Mechanisms

ADS-OT uses six classes of adaptive mechanism. *Adaptive monitoring* changes logging depth, capture, sensitivity, or analytic priority. *Adaptive access control* adds step-up authentication, session recording, or stricter vendor-access conditions. *Adaptive deception* activates decoy services, honey credentials, fake project files, or isolated targets. *Adaptive segmentation* tightens conduits or isolates non-critical paths. *Adaptive protocol validation* scrutinises write commands, firmware operations, and unusual protocol sequences. *Adaptive response readiness* stages playbooks, notifies operators, verifies recovery, and raises analyst priority.

### E. Governance Feedback

This layer makes adaptation safe — safe-action catalogue, safety shield, approval rules, audit, rollback, post-action review; without it, an adaptive system is just another source of operational uncertainty.

## V. A Lightweight POMDP Formulation

Acting on incomplete, noisy evidence, we model defensive adaptation as a POMDP:

$$M = (S, A, O, T, Z, R, \gamma), \tag{2}$$

where $S$ is the hidden state space, $A$ the action space, $O$ the observation space, $T$ the state-transition model, $Z$ the observation model, $R$ the reward function, and $\gamma$ the discount factor [8]. Reinforcement learning seeks a policy $\pi(b)$ mapping a belief state $b$ to an action [7]; after action $a_t$ yields observation $o_{t+1}$, the belief updates as $b_{t+1} = \tau\,(b_t, a_t, o_{t+1})$ by Bayesian or particle filtering. Figure 2 traces the loop.

Notation: uppercase symbols denote spaces or functions; subscripted lowercase symbols their realisations at step $t$. Thus $s_t \in S$ is the hidden state ($s_t$, equivalently $x_t$, in Figure 2), $a_t \in A$ the executed action, $o_{t+1} \in O$ the observation, $T\,(s_{t+1} \mid s_t, a_t)$ and $Z(o_{t+1} \mid s_{t+1}, a_t)$ the transition and observation kernels, $r_t = R(s_t, a_t)$ the per-step reward, and $\gamma \in [0, 1)$ the discount. In the SME case study, one component of $s_t$ is the engineering workstation's compromise status, a typical $a_t$ "require step-up authentication at the vendor gateway", a typical $o_{t+1}$ an out-of-hours remote-access alert. Likewise $b_t$ is the belief, $\pi$ the policy, $\tau$ the belief update, $\Sigma$ the shield of (4), $B$ the burden index of (5).

**Hidden states** span assets normal, exposed, suspicious, compromised, or recovering; attacker stage; AI capability level; production mode; defensive posture; asset criticality; and cyber-physical impact potential.

**Actions** are controlled adaptations: adjust zone monitoring, require stronger authentication, enable session recording, activate deception, restrict high-risk write commands outside approved windows, raise protocol validation, tighten a conduit, escalate to approval, or hold posture.

**Observations** include failed logins, unusual remote access, new assets, engineering-workstation and PLC programming events, protocol command sequences, IDS alerts, process deviations, log inconsistencies, threat intelligence, and operator reports.

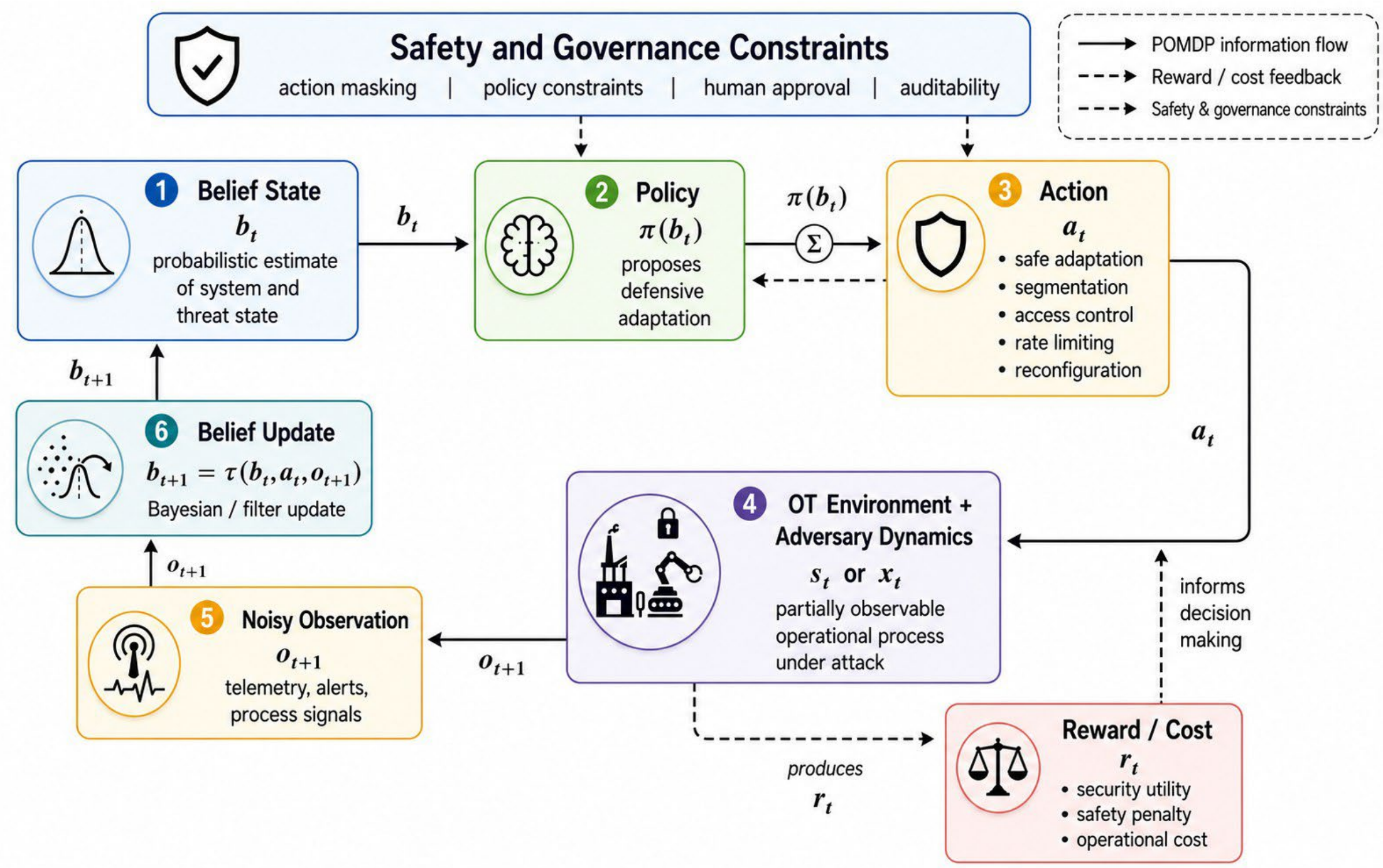

Fig. 2. POMDP loop for ADS-OT. The learned policy *proposes* $\hat{a}_t = \pi(b_t)$; execution authority rests with the safety shield Σ, which gates proposals as in (4). The loop closes through noisy observations and belief update; $r_t$ captures security utility, safety penalties, and operational cost; $s_t$ and $x_t$ both denote the hidden state.

**The realised reward** encodes site priorities:

$$r_t = G_t^{\text{risk}} + G_t^{\text{res}} - C_t^{\text{op}} - C_t^{\text{change}} - C_t^{\text{fp}} - P_t^{\text{safety}}, \quad (3)$$

where $G_t^{\text{risk}}$ is cyber-risk reduction, $G_t^{\text{res}}$ resilience gain, $C_t^{\text{op}}$ operational cost, $C_t^{\text{change}}$ discourages excessive adaptation, $C_t^{\text{fp}}$ penalises false positives, and $P_t^{\text{safety}}$ is a strong safety penalty that only *shapes* learning during training — hard safety is enforced separately and unconditionally by the shield in (4).

**The safety shield** is independent of the learned policy: before execution, $\Sigma(a_t, b_t)$ checks plant constraints, change-control rules, approvals, and rollback feasibility. The executed action is therefore not $\pi(b_t)$ but

$$a_t^{\text{exec}} = \begin{cases} \pi(b_t) & \text{if } \Sigma(\pi(b_t), b_t) = 1, \\ a^{\text{safe}} & \text{otherwise,} \end{cases} \quad (4)$$

where $a^{\text{safe}}$ is a designated fallback such as "hold posture" or "increase passive monitoring only". A hard mask matters in OT: a reward shaped to discourage unsafe actions is not a circuit that prevents them.

## VI. From Controls to Adaptive Actions

ADS-OT should not invent arbitrary actions: adaptive actions operationalise risk-based controls the site already trusts. The workflow: document OT risks from incident reports, threat intelligence, advisories, and assessments; map them to ATT&CK Enterprise and ICS techniques; map techniques to controls; classify each control as foundational or adaptive; convert adaptive controls into safe POMDP actions; and train and evaluate the policy in a cyber range or digital twin. The result is end-to-end traceability from risk to action — something operators, auditors, and incident commanders demand the moment something unusual happens.

Table II maps seven control areas; the actions are familiar, but their intensity and timing are now selected dynamically.

## VII. A Plant-Floor Scenario

Consider a manufacturing site with three production cells, several PLCs, an engineering workstation, a security-monitoring server, IIoT sensors, and a vendor remote-access gateway. It has segmentation, role-based access control, backups, central logging, and a remote-access procedure, but cannot run deep inspection everywhere (latency) or demand high-friction authentication for every action (time-sensitive maintenance).

The attack begins quietly: public sources suggest the likely remote-access vendor and engineering software; generative tools produce a maintenance-themed message naming a plausible firmware update and a real ticket format; one engineer, mid-shift, enters credentials into a spoofed portal.

The first observations are weak: an unusual login location, a session outside normal hours, command timing that does not match maintenance. A static system would surface isolated alerts; ADS-OT instead updates its belief, and the probability of credential compromise climbs. The policy proposes low-risk adaptations: step-up authentication for new vendor sessions, session recording at the gateway, deeper monitoring

TABLE II
TRANSLATING RISK-BASED CONTROLS INTO ADAPTIVE DYNAMIC SECURITY ACTIONS

| **Control area** | **AI-powered OT risk** | **Adaptive action** |
|---|---|---|
| Continuous monitoring | AI-assisted reconnaissance of engineering assets. | Increase logging, packet capture, and analytic priority for engineering traffic. |
| Access enforcement | Credential theft via AI-generated maintenance phishing. | Require step-up authentication, session recording, and temporary approval for vendor access. |
| Boundary protection | Lateral movement from enterprise IT to OT. | Tighten selected conduits and raise gateway inspection for a defined period. |
| Configuration management | Unauthorised PLC logic change outside maintenance. | Increase validation of write commands and compare against approved change tickets. |
| Deception | Attacker probing for engineering project files. | Activate decoy project files, honey credentials, and isolated deception endpoints. |
| Incident response | Ambiguous signs of compromise during production. | Stage playbooks, notify OT operators, and raise analyst priority without interrupting control loops. |
| Recovery | Suspected destructive or ransomware-like action. | Verify backup readiness, recovery paths, and security log integrity before stronger containment. |

of engineering-workstation traffic. None of these touches the process.

Hours later, engineering-software activity appears from a workstation not scheduled for maintenance; belief in compromise rises. The policy proposes decoy project files, increased validation of PLC write commands, and analyst notification. Production continues; no PLC address, safety logic, or asset changes.

If the activity proves to be legitimate maintenance, the policy reduces friction and returns to baseline; if evidence sharpens, it proposes stronger actions — temporary write-command restriction, segmentation of a non-critical pathway — subject to operator approval. Belief-state adaptation acts earlier than static thresholds while avoiding unnecessary disruption.

## VIII. SME MANUFACTURING CASE STUDY

We evaluated the framework on an illustrative SME precision-manufacturing scenario inside an OT cyber-range simulation. The site has 120 employees and three production cells; its 74-asset OT estate includes 12 PLCs, a SCADA server, a security-monitoring server, an engineering workstation, two remote-access gateways, 18 IIoT sensing or edge devices, and supporting switches, HMIs, and operator workstations. Figure 3 depicts the emulated topology — zones and conduits per IEC 62443-3-2, levels per the Purdue model in NIST SP 800-82r3 — with campaign path and posture actions overlaid. The threat scenario is the AI-assisted campaign described above — reconnaissance, phishing, credential theft, remote session, lateral movement, attempted unauthorised PLC write — matching recognised ATT&CK for ICS behaviours (spearphishing attachment, valid accounts, remote services, unauthorised program download).

We compared four strategies under identical conditions: a *static baseline* (existing segmentation, logging, access control); *static risk-based controls* (enhanced monitoring at gateway and engineering workstation); *rule-based adaptation* (threshold-triggered); and ADS-OT (a POMDP/RL policy adapting monitoring, authentication, deception, protocol validation, and response readiness from the belief state). All respected the same hard safety constraints: no active scanning of fragile PLCs, no safety-logic modification, no blocking of critical flows without approval.

*Metric definitions. Attack completion* is the fraction of episodes where the attacker reaches process impact. *MTTD* (*MTTC*) is the mean time from the first attacker action to first correct detection (to containment, when the attack can no longer progress), in simulated minutes, over episodes where the event occurs. *False positives per day* counts benign events escalated for analyst review. The *burden index* is defined in (5) below.

Across 1,000 attack episodes, each kill-chain stage either progressed, was detected and contained, or stalled because posture made the next step too costly. Figure 4(a) shows attack-path survival. The static baseline leaves a 24.7% probability of reaching process impact; static risk-based controls reduce it to 13.1%; rule-based adaptation to 7.1%, at higher burden because thresholds trigger more false positives; ADS-OT reaches 3.3% by adapting earlier, when weak signals about credential misuse and workstation behaviour begin to correlate.

Table III consolidates the metrics: versus the static baseline, ADS-OT cuts attack completion by 86.6%, mean time to detection (MTTD) by 74.0%, and mean time to containment (MTTC) by 71.0%. Rule-based adaptation also lowers completion, but with markedly higher false positives and burden — a familiar OT trade-off, thresholds being effective but noisy. ADS-OT's burden exceeds static risk-based controls, an overhead justified by far stronger risk reduction.

The burden index aggregates analyst workload $\tilde{L}_a$, storage $\tilde{L}_s$, compute $\tilde{L}_c$, and added control-plane latency $\tilde{L}_\ell$, normalised, as

$$B = 100 \sum_{i\in\{a,s,c,\ell\}} w_i \tilde{L}_i, \qquad \sum_i w_i = 1, \tag{5}$$

where each $\tilde{L}_i \in [0, 1]$ is normalised against the rule-based strategy at full alert volume (anchoring the scale) and the weights $w_i$ reflect site priorities.

Figure 4(b) reframes the same data on a risk–cost plane. ADS-OT's burden of 36 exceeds the static baseline (22) and static risk-based controls (31) because adaptation has a price: deeper logging and packet capture consume storage and compute, deception requires upkeep, the belief estimator adds processing. The comparison that matters is with rule-based adaptation (44): escalating from a fused belief rather than per-signal thresholds triggers far less often on benign activity, roughly halving false positives (11 versus 22 per day) while

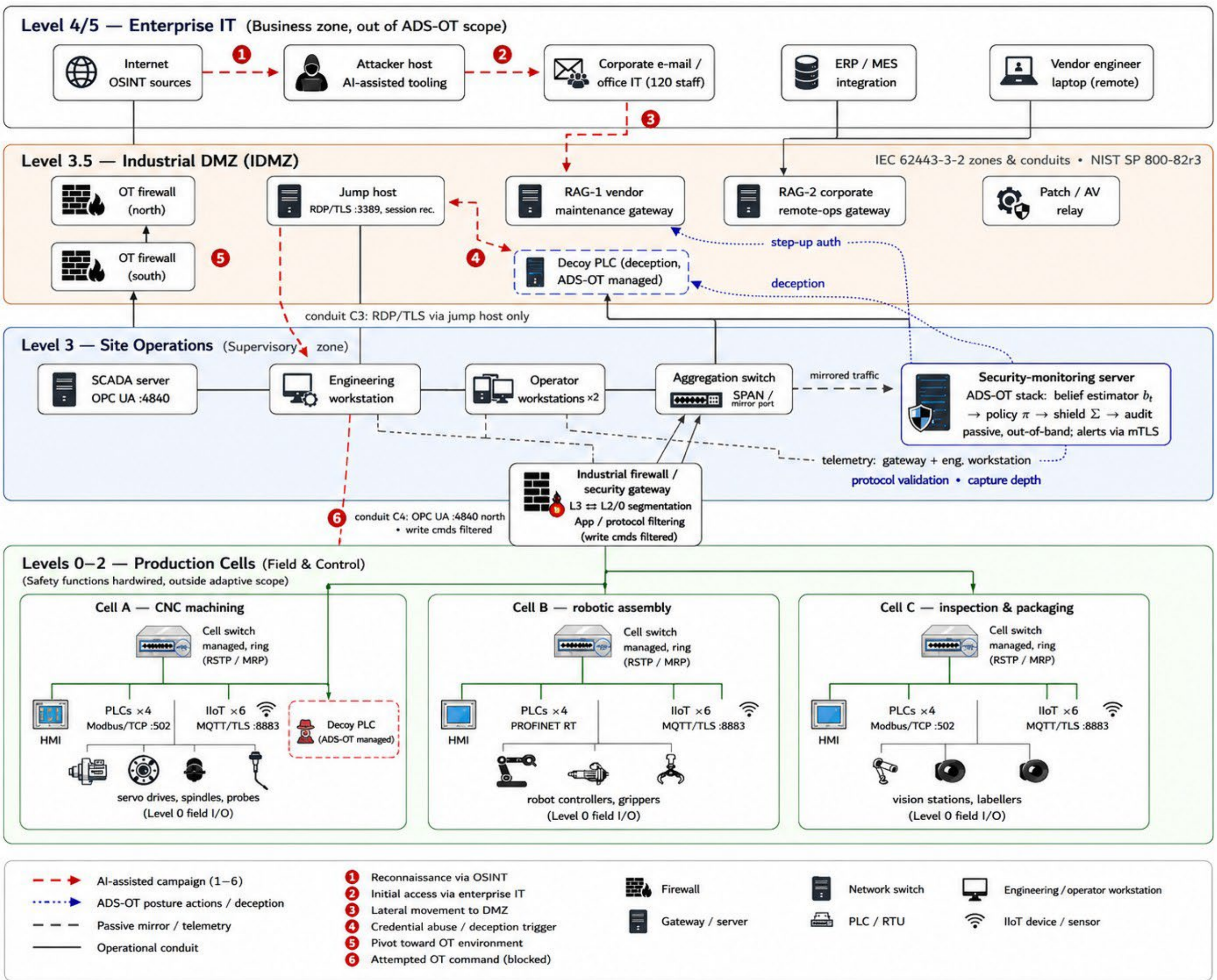

Fig. 3. Case-study testbed: the emulated 74-asset SME plant of Section VIII across three production cells, safety functions hardwired and outside the adaptive scope.

TABLE III
CASE STUDY RESULTS ACROSS 1,000 SIMULATED ATTACK EPISODES (SME MANUFACTURING SITE)

| **Strategy** | **Attack completion** (%) | **Δ vs. baseline** (%) | **MTTD** (min) | **MTTC** (min) | **False positives** (per day) | **Burden index** (0–100) |
|---|---|---|---|---|---|---|
| Static baseline | 24.7 | — | 196 | 310 | 9 | 22 |
| Static risk-based controls | 13.1 | -47.0 | 134 | 220 | 13 | 31 |
| Rule-based adaptation | 7.1 | -71.3 | 86 | 145 | 22 | 44 |
| **ADS-OT** | **3.3** | **-86.6** | **51** | **90** | **11** | **36** |

*Lower is better in all columns.* Burden index: Eq. (5).

delivering the largest completion reduction (86.6%). This trade-off decides adoption: operators rarely reject adaptive defence as ineffective; they reject it as noisy. ADS-OT earns its place by being both effective and tolerable.

The simulation assumes reliable gateway and workstation telemetry and a policy trained on representative scenarios; a less mature SME would run ADS-OT as advisory, under closer review. The takeaway is the evaluation pattern: judge adaptive OT defence on whether it reduces attack progression, detection, and containment times without excessive false positives, friction, or unsafe interventions.

## IX. GOVERNANCE AND SAFETY

Adaptive OT security must be governed as decision support and controlled automation, not autonomous defence replacing human judgement. Four elements anchor that framing.

**A safe-action catalogue.** Every action is defined before deployment: assets touched, cyber benefit, operational side effect, required approval, maximum duration, rollback procedure, audit fields. "Increase passive logging on the Cell 2 gateway for 30 minutes" is a well-defined action; "block suspicious traffic" is not, unless protocol, endpoints, duration, and approval path are specified.

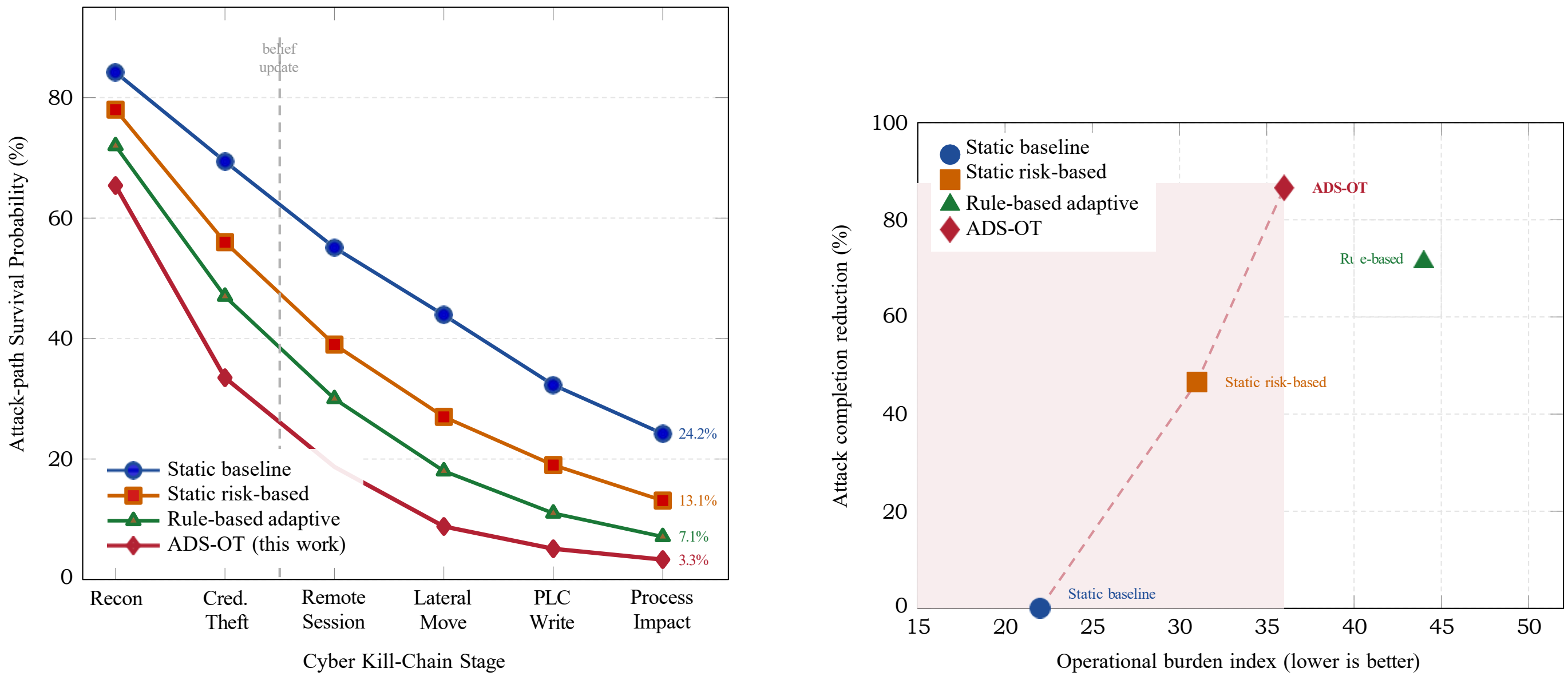

(a)

(b)

Fig. 4. (a) Attack-path survival: ADS-OT pulls ahead between credential theft and remote-session anomalies, where belief fusion first triggers a posture change. (b) Risk–cost trade-off: ADS-OT beats rule-based adaptation on both axes and dominates the static baselines.

**An independent safety shield.** The shield blocks actions violating hard constraints: no active scanning of fragile PLCs; no unauthorised changes to safety instrumented systems; no segmentation change during emergencies unless pre-approved; no write-command restriction during safety-critical recovery; no action exceeding control-traffic latency thresholds. It must be independently verifiable and independent of the learned policy: if the policy is wrong about a constraint, the shield catches it.

**Explainability and audit.** Operators must understand why a recommendation was made: dominant belief factors, asset criticality, expected risk reduction and cost, safety rationale, rollback condition. Every executed action produces an audit record: time, belief summary, action, safety checks, approvals, result, post-action observations, rollback status.

**Human responsibility.** Responsibility cannot be delegated to a model: a RACI matrix specifies who is responsible, accountable, consulted, and informed for each action class. Low-risk passive adaptations can eventually be automated; high-impact actions remain approval-based, especially where safety-critical.

## X. Evaluation Metrics for Practitioners

ADS-OT must be assessed on cyber, operational, and governance criteria together: detecting more attacks while disrupting production is unacceptable; never adapting is equally inadequate. Table IV summarises the metric families used in the case study, recommended for any cyber-range comparison.

## XI. Lessons for Practice

Reinforcement learning cannot replace OT security engineering judgement; five lessons should guide any real-world implementation.

**Do not treat RL as a magic optimiser.** RL learns sequential decisions under uncertainty; it is not intelligent in any general sense. Action space, reward, constraints, and simulation environment determine what a policy can learn; poor design produces brittle or unsafe policies regardless of model size.

**Start with controls you already trust.** Adaptive actions should come from familiar controls and procedures; ADS-OT changes how they are prioritised and adjusted, not the requirement to govern them.

**Train offline, deploy conservatively.** Learning belongs in a digital twin, cyber range, or emulation; live OT systems are not exploration platforms. Deployment starts advisory; automation begins with low-risk passive actions such as increased logging.

**Make safety a hard boundary.** A reward penalty is not enough: hard shield constraints block unsafe changes, and the shield should be simpler, more transparent, and more conservative than the policy.

**Measure operational cost.** Improvements that create false positives, fatigue, storage cost, latency, or mistrust will not survive production; burden is part of the evaluation, not an afterthought.

## XII. Implementation Roadmap

Deployment is staged; each stage builds confidence before any autonomous action is allowed.

*Stage 1: Baseline the environment.* Engineering hygiene first: asset inventory, zone and conduit model, account inventory, remote-access register, communication baseline, safety constraints, critical-process list. Without this foundation, an adaptive policy optimises over an unreliable view of the plant.

TABLE IV
EVALUATION METRICS FOR ADS-OT

| Category | Example metric | Purpose |
|---|---|---|
| Cyber-risk reduction | Reduction in attack-path completion probability. | Measures defensive effectiveness. |
| AI-threat resilience | Disruption of AI-assisted phishing, reconnaissance, evasion, and chaining scenarios. | Tests relevance to the threat model. |
| Detection | MTTD; ATT&CK technique visibility. | Measures situational awareness. |
| Response | MTTC; successful rollback rate. | Measures response quality. |
| OT safety | Safety-constraint violations; latency added to critical flows. | Tests control-environment suitability. |
| Operational burden | False positives; analyst workload; storage and compute overhead. | Measures sustainability. |
| Adaptation quality | Risk reduction per adaptation; unnecessary-adaptation rate. | Measures policy efficiency. |
| Governance | Fraction of actions with explanation, approval, and audit record. | Measures assurance and accountability. |

*Stage 2: Identify adaptive candidates.* Not every control should adapt; some define the security foundation and stay fixed. Good early candidates are low-risk, measurable, reversible: passive logging, packet capture, session recording, analyst prioritisation, deception. Disruptive actions — segmentation tightening, command restriction — come only after extensive review.

*Stage 3: Build the training environment.* Train and stress-test the policy in an OT cyber range, emulation, or digital twin containing benign maintenance, AI-assisted social engineering, remote-access compromise, lateral movement, process anomalies, false positives, and delayed telemetry, with realistic operational costs: a simulation that ignores workload or latency teaches the wrong behaviour.

*Stage 4: Run in advisory mode.* The system recommends, explains, and records operator accept/reject decisions — valuable even with nothing automated, revealing whether the belief model is credible, recommendations understandable, and actions matched to practice.

*Stage 5: Automate only the lowest-risk actions.* Automation begins with passive, reversible actions unlikely to affect production: time-limited logging increases, packet capture, deception in isolated locations, analyst priority. High-impact actions remain approval-based unless pre-authorised for a specific emergency.

## XIII. RESEARCH CHALLENGES

Several challenges remain open. State spaces are large, so practical models need abstraction by zone, function, criticality, or attack relevance — concretely, a *factored* belief (per-zone compromise likelihoods rather than a joint distribution over every asset) maintained by discrete Bayes or particle filtering, with hierarchical actions keeping the action space tractable as the estate grows. Simulation realism is difficult: a policy trained on unrealistic attack paths or simplified dynamics may fail in production, so scenarios need grounding in testbed data, threat intelligence, operator review, and incident exercises. Training must happen entirely offline, with the shield of (4) active during exploration as well as deployment — the learner must never discover unsafe actions on the live plant — and transfer follows the staged, monitor-first rollout above, the policy treated as a proposal source until it has earned operator trust. Attackers may adapt to the adaptive defence itself, motivating randomisation, adversarial testing, and policy rotation, each safety-constrained. Reward design is socio-technical: security, engineering, safety, and operations value risk, availability, and intervention cost differently, so the reward function must be negotiated and reviewed. Data quality is often limited: ADS-OT should degrade gracefully and improve incrementally rather than assume perfect telemetry. Finally, assurance is unresolved: regulators, insurers, and operators will require evidence that adaptive decisions are safe, explainable, and reversible — audit trails, testing protocols, possibly certification.

## CONCLUSION

AI-POWERED attacks challenge the assumption that OT security can remain static between assessments: they accelerate reconnaissance, frame social engineering in the operator's language, support evasion, and chain weaknesses across IT/OT boundaries. Defence must become adaptive, yet stay constrained by safety, reliability, and governance. ADS-OT models defensive adaptation as a POMDP, learns policies by RL under uncertainty, and constrains execution through safe-action catalogues, safety shields, operator approval, audit, and rollback. Its mechanisms are familiar; what changes is their timing, intensity, and prioritisation. In a 1,000-episode SME case study, this reduced attack-path completion by 86.6%, MTTD by 74.0%, and MTTC by 71.0% versus a static baseline, with lower operator burden than threshold-driven adaptation. The next step is validation in live OT cyber ranges and digital twins against the metrics defined here. The goal is not autonomous control of industrial processes; it is security that remains adaptive when attackers become adaptive.

## ACKNOWLEDGMENT

The author thanks former colleagues and industrial cybersecurity managers for discussions on OT threat models and operational realities.

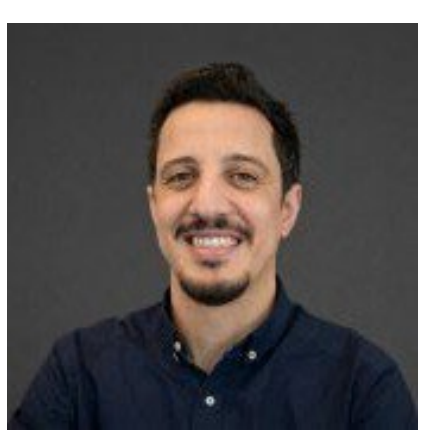

**Mohamed Chahine Ghanem** (Member, IEEE) is an Associate Professor in Cyber Security in the School of Computer Science and Mathematics at Keele University, UK, and Co-founder of CyDRA Limited. He previously spent over 15 years in senior law-enforcement and corporate positions as a principal cyber investigator and risk auditor. He holds an Engineering degree in Computer Systems, an MSc in Digital Forensics & IT Security, and a PhD in Cyber Security Engineering from City, University of London; he is a Senior Fellow of the HEA and holds CISSP, CPCI, and multiple GIAC certifications. He leads research projects on applied AI and cyber security and has published numerous papers in Q1/Q2 journals. He is Editor-in-Chief of the Journal of Cyber Security and Risk Auditing and an Associate Editor of ACM Digital Threats: Research and Practice.